\documentclass{ws-ijmpd}

\usepackage{epsf}

\begin{document}

\hyphenation{ge-ne-ra-li-ty}
\hyphenation{cos-mo-lo-gi-cal}
\hyphenation{ex-po-nen-ti-al}
\hyphenation{so-lu-ti-ons}
\hyphenation{co-ef-fi-ci-ents}
\hyphenation{ex-po-nen-ti-al-ly}
\hyphenation{cor-res-pon-ding}
\hyphenation{as-sump-ti-ons}
\hyphenation{equa-ti-ons}

\def\nocropmarks{\vskip5pt\phantom{cropmarks}}

\markboth{G.V. Vereshchagin}
{On stability of simplest nonsingular inflationary cosmological models...}

\catchline{}{}{}{}{}

\title{On stability of simplest nonsingular inflationary cosmological models within general relativity and gauge theories of gravity}

\author{G.~V. VERESHCHAGIN}

\address{Belorussian State Pedagogical University, Physics Department, \\
Sovetskaya str. 18, 220050 Minsk, Belarus \\
and \\
I.C.R.A. - International Center for Relativistic Astrophysics, \\
University of Rome "La Sapienza", Physics Department, \\
P.le A. Moro 5, 00185 Rome, Italy. \\
Email: veresh@icra.it}

\maketitle

\begin{abstract}
In this paper we provide approximate analytical analysis of stability of nonsingular inflationary chaotic-type cosmological models. Initial conditions for nonsingular solutions at the bounce correspond to dominance of potential part of the energy density of the scalar field over its kinetic part both within general relativity and gauge theories of gravity. Moreover, scalar field at the bounce exceeds the planckian value and on expansion stage these models correspond to chaotic inflation. Such solutions can be well approximated by explicitly solvable model with constant effective potential (cosmological term) and massless scalar field during the bounce and on stages of quasi-exponential contraction and expansion. Perturbative analysis shows that nonsingular inflationary solutions are exponentially unstable during contraction stage. This result is compared with numerical calculations.
\end{abstract}

{\it Keywords: initial singularity, inflation, stability, gauge theories of gravity}

\vspace{5mm}

PACS: 04.20.Dw, 04.50, 98.80.Cq

\vspace{5mm}


\section{Introduction}

In conventional cosmological scenario the Universe has a beginning in time, so called Big Bang, associated with initial singularity of Friedmann equations \cite{Kol90}. Singular state with infinite values of energy density and curvature is unacceptable from physical point of view and possibly marks the edge of applicability of General Relativity (GR). In spite of some progress in last decades, attempts to resolve this problem within GR had no success \cite{Linde,BMS}.

Correct description of the early Universe when the energy density was very high should be given by particle physics \cite{Linde}. At the same time experimental and observational data are still insufficient to discriminate between different models of elementary particles physics. However in most models scalar fields play an important role. One of the main discovery in modern cosmology is a possibility to explain a number of puzzles of Big Bang model by inclusion of short stage of exponential expansion of the Universe, so called inflation \cite{Linde,Olive}. During this stage the energy density of the Universe is dominated by a scalar field (or fields) and transition to usual hot Universe is achieved by creation of relativistic particles at the end of inflation \cite{Reheating}.

Unfortunately even within inflationary models there is little chance to avoid cosmological singularity \cite{Bor01}. The evolution of the Universe is usually considered from the planckian moment when description in terms of classical theory of gravitation becomes possible. At contraction stage the singular state is unavoidable in most models \cite{BMS,BGZH}.

Nevertheless, some nonsingular inflationary solutions exist \cite{BGZH,Kam97,mv1,Fak00}. Although generality of these solutions is known to be small \cite{Page84}, they are still of some interest in the literature \cite{Fak00,Mink02,Mink02a,Mink03a}. In fact, inflationary initial conditions for the scalar field are the same for bouncing inflationary models. It is known that in order to have nonsingular cosmological model with the scalar field the potential part of the energy density at the bounce must be larger than the kinetic part. These conditions lead automatically to subsequent inflation in the case when the scalar field exceeds the planckian value at the bounce. Therefore it was expected that nonsingular inflationary models are also quite probable \cite{mv1,Mink02}.

Generality of inflation was studied, for instance, in \cite{BGZH,Kam97}. In \cite{BGZH} it was pointed out that although infinite number of nonsingular inflationary solutions exist within GR, exponential contraction stage of these solutions is unstable. By defining a measure on the phase plane where the Hubble parameter equals zero, the probability of bouncing inflationary solutions was estimated to be of order unity \cite{Mink02,Mink02a,Mink03a}. At the same time, probability of nonsingular solutions with massive scalar field was estimated \cite{Star} to be $P\sim 10^{-43}$.

From the other hand, deviations from Riemannian geometry can occur at very high energy density. In more general spacetimes GR should be replaced by gauge theories of gravity (GTG) that are natural generalizations of GR \cite{Hehl}. It was shown that nonsingular cosmological solutions within GTG exist in the case of spinning matter \cite{Hehl}, usual matter \cite{Acta98} and matter in the form of scalar field \cite{mv1}. Moreover, when energy density becomes small enough the deviation from GR becomes negligible and at late stages the Universe can be described by Friedmann equations. This could allow generalizing standard cosmology and resolving the problem of initial singularity.

In order to analyze the generality of initial conditions for nonsingular cosmological models within GR as well as within GTG in this paper perturbations for nonsingular solutions are studied. The paper is organized as follows. In the Sec.2 we discuss bouncing inflationary models within GR in the case of minimal coupling between the scalar field and gravitation. In the Sec.3 coupling is assumed to be nonminimal within the same models. In the Sec.4 bouncing inflationary models within GTG are discussed. In Sec.5 and 6 perturbations are studied analytically. In Sec.7 we present numerical results and comparison with analytical calculations. The last two sections contain discussion and conclusions.

\section{Bounce in inflationary models within GR}

Cosmological Fridmann equations for homogeneous isotropic models in the case of single real scalar field $\varphi$ minimally coupled to gravity read

\begin{equation}
\begin{array}{rcl}
\displaystyle{\frac{k}{a^2}+\left(\frac{\dot a}{a}\right)^2}&=&\displaystyle{\frac{8\pi}{3M_p^2}\left[\frac{1}{2}\dot\varphi^2+V(\varphi)\right]}, \\
\\
\displaystyle{\frac{\ddot a}{a}}&=&\displaystyle{-\frac{8\pi}{3M_p^2}\left[\dot\varphi^2-V(\varphi)\right]}.
\label{FrEqns}
\end{array}
\end{equation}
where $V(\varphi)$ is effective potential of the scalar field, $a(t)$ is the scale factor of the Universe, $M_p$ is planckian mass, $k$ corresponds to the sign of space curvature, and the dot denotes time derivative\footnote{$\hbar=c=1$ system is used.}.

Conservation law follows from (\ref{FrEqns}) as

\begin{equation}
\label{cl}
\ddot{\varphi}+3\left(\frac{\dot a}{a}\right)\dot\varphi+\frac{\partial V}{\partial\varphi}=0.
\end{equation}

From (\ref{FrEqns}) it also follows \cite{Mink02} that nonsingular solutions exist for the Universe with positive curvature ($k=+1$) for initial conditions at the bounce\footnote{It is convenient to measure time from the moment of bounce.} ($\dot a(0)=0$, $\ddot a(0)>0$) as

\begin{equation}
\label{GRbounce}
\dot\varphi^2(0)<V(\varphi(0)).
\end{equation}

Solution of (\ref{FrEqns}) in WKB approximation \cite{Star} and qualitative analysis of dynamical system \cite{BGZH} for the case of massive scalar field ($V_1=m^2\varphi^2/2$ with $m\sim 10^{-6}M_p$) show, that nonsingular solutions of this kind have practically zero measure.

The effective equation of state for most solutions approach `stiff' equation $p=\rho$, where $\rho$ is the energy density and $p$ is the pressure within GR \cite{BGZH} and ultrarelativistic equation $p=\rho/3$ within GTG \cite{Ver03} on contraction stage.

It is interesting to note that if initial conditions for the scalar field satisfy conditions (\ref{GRbounce}) and $\varphi(0)>M_p$, then nonsingular solutions correspond to chaotic inflation \cite{Lin83} on expansion stage. The latter begins right after the transition from contraction to expansion.

Inflationary stage in chaotic scenario lasts until the scalar field reaches planckian value $\varphi\simeq M_p$. From that moment scalar field starts to oscillate with damped amplitude within GR. In GTG oscillations in scalar field can be accompanied by oscillations of the Hubble parameter \cite{Mink03b}. Usually it is assumed that during coherent oscillations of the scalar field the energy density of the Universe is transferred into the form of ultrarelativistic matter.

\section{Nonminimal coupling}

In the previous section the scalar field assumed minimally coupled to gravitation. However, it is necessary for the renormalizability of the theory \cite{Fre74,Fre74a,Lin82a} that the lagrangian contains an explicit coupling term between the scalar field and the scalar curvature. The correct form of the lagrangian appears to be

\begin{equation}
L=\frac{M_p^2 R}{16\pi}+\frac{1}{2}\xi\varphi^2 R+\frac{1}{2}\varphi_{;\mu}\varphi^{;\mu}-V(\varphi),
\label{Lnc}
\end{equation}
where the semicolon denotes covariant derivative.

The conformal coupling case corresponds to $\xi=-1/6$. For arbitrary $\xi$ cosmological equations are \cite{Fak00}

\begin{equation}
\begin{array}{l}
\displaystyle{3\left[\left(\frac{\dot a}{a}\right)^2+\frac{k}{a^2}\right]\left(\frac{M_p^2}{8\pi}+\xi\varphi^2\right)=\frac{1}{2}\dot\varphi^2+V(\varphi)-6\xi\left(\frac{\dot a}{a}\right)\varphi\dot\varphi}, \\ \\
\displaystyle{\left[\frac{\ddot a}{a}-\left(\frac{\dot a}{a}\right)^2-\frac{k}{a^2}\right]\left(\frac{M_p^2}{8\pi}+\xi\varphi^2\right)=-\frac{1}{2}\dot\varphi^2+\xi\left(\frac{\dot a}{a}\right)\varphi\dot\varphi-\xi\dot\varphi^2-\xi\varphi\ddot\varphi}.
\end{array}
\label{cce}
\end{equation}

The conservation law (\ref{cl}) now generalized as

\begin{equation}
\label{clc}
\begin{array}{l}
\displaystyle{\ddot{\varphi}+3\left(\frac{\dot a}{a}\right)\dot\varphi-6\xi\left[\frac{\ddot a}{a}+\left(\frac{\dot a}{a}\right)^2+\frac{k}{a^2}\right]\varphi+\frac{\partial V}{\partial\varphi}=0}.
\end{array}
\end{equation}

Nonsingular cosmological model was obtained in \cite{Fak00} based on equations (\ref{cce}),(\ref{clc}). It was shown, that during expansion this model corresponds to chaotic inflation \cite{Lin83} in spite of the fact, that $\varphi(0)<M_p$.

Notice, that both on quasi-exponential stages of contraction and expansion as well as during transition from contraction to expansion the scalar field changes slowly. This corresponds to the following assumption

\begin{equation}
\varphi=\varphi_0=\mbox{const},
\label{ficonst}
\end{equation}
and as a result

\begin{equation}
\begin{array}{c}
\ddot\varphi=\dot\varphi=0, \\ \\
\frac{\partial V}{\partial\varphi}=\frac{\partial V}{\partial t}=0, \\ \\
V=V_0=\mbox{const}.
\end{array}
\label{ficonseq}
\end{equation}

In this approximation equations (\ref{cce}) reduce to

\begin{equation}
\begin{array}{rcl}
\displaystyle{\frac{\ddot a}{a}}&=&\displaystyle{\frac{8\pi V_0}{3M_p^2(1+\frac{8\pi}{M_p^2}\xi\varphi_0^2)}}.
\label{transeqc}
\end{array}
\end{equation}

This is the second Friedmann equation with modified `newtonian' constant and with the same assumptions. Therefore all results obtained for nonsingular models described by equations (\ref{FrEqns}) in approximation (\ref{ficonst}),(\ref{ficonseq}) are valid for nonsingular models with nonminimal coupling.

\section{Bounce in inflationary models within GTG}

Cosmological equations of GTG in the case of single real scalar field with the effective potential $V(\varphi)$ in the homogeneous isotropic Universe are \cite{Mink02,Ver03}

\begin{equation}
\label{gcfe2}
\begin{array}{l}
\begin{displaystyle}
\frac{k}{a^2}+\left[\left(\frac{\dot a}{a}\right)\left(1-3\beta\dot{\varphi}^2 Z^{-1}\right)-
3\beta V' \dot{\varphi} Z^{-1}\right]^2=
\end{displaystyle} \\
\quad\quad\quad\quad\quad\quad\quad
\quad\quad\quad\quad\quad\quad\quad
\begin{displaystyle}
\frac{8\pi }{3M_{p}^{2}}\,
\left[\frac{1}{2}\dot{\varphi}^2+V-\frac{\beta}{4}\left(4V-
\dot{\varphi}^2\right)^2\right]\,Z^{-1},
\end{displaystyle}
\end{array}
\end{equation}

\vskip4mm

\begin{equation}
\label{Heqn2}
\begin{array}{l}
\displaystyle{\left[\frac{\ddot a}{a}-\left(\frac{\dot a}{a}\right)^2\right]\left(1-3 \beta\dot{\varphi}^2
Z^{-1}\right)+\left(\frac{\dot a}{a}\right)^2\left(1+15\beta\dot{\varphi}^2 Z^{-1}-18\beta^2\dot{\varphi}^4
Z^{-2}\right)+} \\
\displaystyle{12\beta\dot{\varphi}V'Z^{-1}
\left(1-3\beta\dot{\varphi}^2Z^{-1}\right)\left(\frac{\dot a}{a}\right)-3\beta Z^{-1}\left(V''\dot{\varphi}^2-(V')^2+
6\beta\dot{\varphi}^2(V')^2Z^{-1}\right)=} \\
\quad\quad\quad\quad\quad\quad\quad
\quad\quad\quad\quad\quad\quad\quad
\quad\quad\quad
\displaystyle{\frac{8\pi}{3M_{p}^{2}}\left[V-\dot{\varphi}^2-\frac{\beta}{4}\left(4V
-\dot{\varphi}^2\right)^2\right]\,Z^{-1},}
\end{array}
\end{equation}

\vskip4mm

where $Z(t)=1+\beta\left[\dot{\varphi}^2(t)-4V(\varphi(t))\right]$,
$V'=\frac{dV(\varphi)}{d\varphi}$, $V''=\frac{d^2V(\varphi)}{d\varphi^2}$, indefinite parameter $\beta$ has dimension of inverse energy density and when $\rho\simeq\beta^{-1}$ solutions of (\ref{gcfe2}),(\ref{Heqn2}) deviate from Friedmann solutions. Conservation law (\ref{cl}) follows from cosmological equations of GTG.

Nonsingular inflationary cosmological model was obtained and discussed in \cite{mv1} based on these equations. Later, other models were discussed in \cite{Mink02,Mink03a,Mink03b}. However, they all correspond to chaotic inflation at expansion stage in spite of the presence of some features after inflation, such as oscillations of the Hubble parameter.

Equations (\ref{gcfe2}),(\ref{Heqn2}) can be essentially simplified if we consider exponential contraction or expansion as well as transition from contraction to expansion. During these stages assumptions (\ref{ficonst}),(\ref{ficonseq}) can be applied so (\ref{gcfe2}),(\ref{Heqn2}) reduce to  similar to (\ref{FrEqns}) equations

\begin{equation}
\begin{array}{rcl}
\displaystyle{\frac{k}{a^2}+\left(\frac{\dot a}{a}\right)^2}&=&\displaystyle{\frac{8\pi V_0}{3M_p^2}}, \\
\\
\displaystyle{\frac{\ddot a}{a}}&=&\displaystyle{\frac{8\pi V_0}{3M_p^2}}.
\label{transeq}
\end{array}
\end{equation}

These equations are identical to (\ref{FrEqns}) with $\dot\varphi=0$ and therefore all results obtained from the analysis of GR equations in approximation (\ref{ficonst}),(\ref{ficonseq}) will be valid in the case of GTG in the same approximation.

\section{Zero order equations}

Let us derive equations governing perturbations dynamics during stages of exponential expansion and contraction and during transition from contraction to expansion. Notice first of all, that our perturbative analysis has nothing to do with analysis of relativistic cosmological perturbations. In effect, we are interested in the following problem: what happens with the solution if the initial conditions at the bounce are slightly modified. Remind that usual analysis of perturbations deals with spatial distribution of perturbations at some moment of time.

In the following we assume that the scalar field has the effective potential $V_1$ with corresponding constraint on the mass of the scalar field, or $V_2=\lambda\varphi^4/4$ with $\lambda=10^{-12}$ necessary to obtain small density perturbations after inflation \cite{Linde}.

If we don't take into account the end of inflation and beginning of quasi-exponential contraction stage, nonsingular solutions can be studied in approximation $V\approx V_0=\mbox{const}$. However, we prefer to keep the last term in (\ref{cl}) as a constant. This means we assume that scalar field derivative is nonzero, but the scalar field changes very slowly. Reasons for this `controversial' from the first glance approximation will become clear later.

Further on, we drop term $\dot\varphi^2$ in (\ref{FrEqns}), but keep the second term in (\ref{cl}). With parameters values used in inflationary cosmology this causes small error comparing to numerical solutions. Moreover, the relative difference between exact and approximated solutions (the error) reaches the maximum during transition from contraction to expansion twice but goes to the asymptotic value during de Sitter regimes. Denoting $\psi\equiv\dot\varphi$ and $A=\left(\frac{8\pi V_0}{3M_p^2}\right)^{1/2}$ we arrive to the system of equations

\begin{equation}
\label{system0}
\left\{
\begin{array}{lll}
\dot b    $=$ A^2 a, \\ $ $ \\
\dot\psi  $=$ \displaystyle{-\eta-3\frac{b}{a}\psi,} \\ $ $ \\
\dot a    $=$ b,
\end{array}
\right.
\end{equation}
where $\eta$ is a small constant with dimension $M_p^3$. It is assumed $\eta=m(2V_0)^{1/2}$ when one deals with $V_1$ and $\eta=\lambda^{-1/4}(4V_0)^{3/2}$ in the case of $V_2$.

Analytical solution for this system exists and reads

\begin{equation}
\label{sol0}
\left\{
\begin{array}{lll}
b_0     $=$ a_{0i} A \sinh(At), \\ $ $ \\
\psi_0  $=$ \displaystyle{-\frac{\eta\mathrm{sech}(At)^3}{12A}\left[9\sinh(At)+\sinh(3At)\right],} \\ $ $ \\
a_0     $=$ a_{0i} \cosh(At),
\end{array}
\right.
\end{equation}
where $a_{0i}$ is the minimal value of the scale factor at the moment $t=0$ and $\psi(0)=0$. The Hubble parameter is $H\equiv\frac{b}{a}$ and the corresponding solution is

\begin{equation}
\label{H0}
H_0=A\tanh(At).
\end{equation}

Expanding (\ref{sol0}) in series at $t=0$ and restricting ourselves to second order we arrive to

\begin{equation}
\label{sol0ser}
\left\{
\begin{array}{lll}
H_0     $=$ A^2 t, \\ $ $ \\
\psi_0  $=$ -V_0 t, \\ $ $ \\
a_0     $=$ a_{0i}(1+\frac{1}{2}A^2 t^2).
\end{array}
\right.
\end{equation}

\begin{figure}[htp]
\begin{center}
\epsfxsize=4.5in
\epsfbox{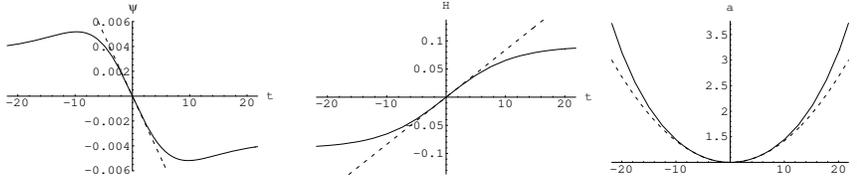}
\end{center}
\caption{Dynamics of cosmological parameters in nonsingular cosmological model with scalar field during transition from exponential contraction stage to inflationary stage is shown. The scalar field potential is chosen to be $V_0=10^{-3}M_p^4$. Dotted curves correspond to approximate solution (\ref{sol0ser}) while firm curves denot exact solution (\ref{sol0}).}
\label{fig1}
\end{figure}
This solution describes transition from contraction to expansion. Both exact and approximate solutions are represented\footnote{All quantities here and below are measured in planckian units.} at fig.\ref{fig1}. One can see, that both Hubble parameter and scalar field derivative $\psi$ are approximately linear in time during transition. Duration of transition from contraction to expansion $t_{tr}$ can be estimated from these expressions since at $t=t_{tr}$ linear regime breaks down. According to (\ref{system0}) it happens when $H=H_{max}=A$ at

\begin{equation}
\label{ttr}
t_{tr}=\frac{2}{A}=\left(\frac{3M_p^2}{2\pi V_0}\right)^{1/2}.
\end{equation}

For typical effective potentials used in inflationary cosmology $V_1$ and $V_2$ the transition time ranges from $1.2t_p$ to $2\cdot10^2 t_p$ for $V_1$ and to $10^5 t_p$ for $V_2$, where $t_p$ is the planckian time.

\section{Perturbation equations}

Assuming standard perturbative scheme

\begin{equation}
\label{pert}
\left\{
\begin{array}{lll}
b     $=$ b_0+b_1, \\ $ $ \\
\psi  $=$ \psi_0+\psi_1, \\ $ $ \\
a     $=$ a_0+a_1,
\end{array}
\right.
\end{equation}
where first order corrections $b_1,\,\psi_1$ and $a_1$ are small, and supposing that functions $b_0,\,\psi_0$ and $a_0$ are given by (\ref{sol0}) the following perturbation equations can be written

\begin{equation}
\label{perteqs}
\left\{
\begin{array}{lll}
\dot b_1    $=$ \displaystyle{\frac{16\pi\eta}{9A M_p^2}T_a(1+2 S_e^2)(a_{0i} C_o+a_1)\psi_1+} \\ $ $ \\ \displaystyle{\frac{8\pi}{27 A^2 M_p^2}\left[9A^2 V_0+\eta^2(4S_e^6-3S_e^2-1)\right]a_1-} \\ $ $ \\
\displaystyle{\frac{8\pi}{3M_p^2}(a_{0i} C_o+a_1)\psi_1^2}, $ $ \\ $ $ \\
\dot\psi_1  $=$ \displaystyle{\frac{1}{A(a_1+a_{0i} C_o)}\left[\eta T_a(2S_e^2+1)(b_1-A T_a a_1)-3A\psi_1(b_1+A a_{0i} S_i)\right]}, \\ $ $ \\
\dot a_1    $=$ b_1,
\end{array}
\right.
\end{equation}
where

\begin{equation}
\label{subs}
\begin{array}{l}
S_i=\sinh[A(t-s)], \quad \quad
C_o=\cosh[A(t-s)], \\ \\
T_a=\tanh[A(t-s)], \quad \quad
S_e=\mathrm{sech}[A(t-s)],
\end{array}
\end{equation}
and $s$ denotes the shift in time before the bounce.

These equations can be integrated only numerically. Even linearized equations where terms like $\psi_1^2$ or $a_1 \psi_1$ are neglected unfortunately are not solvable analytically since their coefficients depend on time\footnote{Full set of equations (\ref{perteqs}) is involved into subsequent numerical analysis since linearized perturbed equations are, in general, equations describing different problem. Numerical calculations of differences of two close solutions agree with the (\ref{perteqs}) solutions but disagree in general with solutions of the linearized system.}.

In order to find approximate equations that represent main properties of solutions of (\ref{perteqs}) we expand time dependent coefficients in (\ref{perteqs}) in series keeping zero order quantities only. This allows to reduce system (\ref{perteqs}) to a system with constant coefficients. However we keep dependence on $s$ and therefore our approximation is valid for every moment of evolution of the model parametrized by $s$, but during short time intervals.

Notice, that usually quantities like $a_1 \psi_1$ and $\psi_1^2$ are neglected since they are second order with respect to first order quantities denoted with subscript `1'. However, in the particular problem we deal with, we focus mainly on perturbations in the quantity $\psi$ and assume that they can be quite large. We will see, that the term with $\psi_1^2$ in (\ref{perteqs}) causes asymmetry in perturbations dynamics during contraction. The term $\psi_1 b_1$ is neglected in the second equation of (\ref{perteqs}) and in denominator of this equation $a_1$ is also neglected.

The new system of equations reads

\begin{equation}
\label{perteqsser}
\left\{
\begin{array}{lll}
\dot b_1      $=$ C_1 a_1+C_2 \psi_1+C_0\psi_1^2, \\ $ $ \\
\dot\psi_1   $=$ C_3 b_1+C_4 a_1+C_5 \psi_1, \\ $ $ \\
\dot a_1     $=$ b_1,
\end{array}
\right.
\end{equation}
where

\begin{equation}
\label{coefs}
\begin{array}{l}
C_0=\displaystyle{-\frac{8\pi a_{0i}}{3M_p^2}\cosh(As)}, \\ \\
C_1=\displaystyle{\frac{8\pi}{27A^2M_p^2}\left[9A^2 V_0+\eta^2(4\mathrm{sech}^6(As)-3\mathrm{sech}^2(As)-1)\right]}, \\ \\
C_2=\displaystyle{-\frac{8\pi a_{0i}\eta}{9A M_p^2}\mathrm{sech}(As)\tanh(As)(5+\cosh(2As))}, \\ \\
C_3=\displaystyle{-\frac{\eta}{A a_{0i}}\mathrm{sech}(As)\tanh(As)(1+2\mathrm{sech}^2(As))}, \\ \\
C_4=\displaystyle{C_3 A\tanh(As)}, \\ \\
C_5=\displaystyle{3A\tanh(As)}.
\end{array}
\end{equation}

Notice now, that the coefficient $C_5$ is always much greater than $C_3$ and $C_4$. In fact, constant $A$ is small since $V\ll M_p^4$. Expanding these coefficients into series and keeping second order terms one can check that $C_5\gg C_3,C_4$. This inequality is valid for all $s$ except for $s=0$ as at that moment $C_3=C_4=C_5=0$. Therefore, equation for $\psi_1$ can be decoupled from (\ref{perteqsser}) and integration gives immediately

\begin{equation}
\label{psi1}
\displaystyle{\psi_1=\delta\psi \exp\left\{C_5 t\right\}=\delta\psi \exp\left\{\frac{2}{M_p}(6\pi V_0)^{1/2}\tanh\left[\frac{2s}{M_p}\left(\frac{2\pi V_0}{3}\right)^{1/2}\right]t\right\}},
\end{equation}
where $\delta\psi$ is initial deviation from solution $\psi_0$.

This constitutes the main result of the paper and shows that for negative $s<0$, i.e. for moments of time after the bounce perturbations of scalar field derivative converge exponentially with time, so the motion of the scalar field is asymptotically stable. On the contrary, on contraction stage with $s>0$ perturbations of scalar field derivative grow exponentially with time and therefore the motion is unstable.

Analytical solutions for functions $b_1$ and $a_1$ in the case $C_0=0$ can also be obtained after substitution of $\psi_1$ from (\ref{psi1}) but they look quite intricate. The most important point concerning these solutions is that both for contraction and expansion stages they can be represented as

\begin{equation}
\label{a1b1app}
a_1=h_1 exp\{\alpha_1 t\}+h_2 exp\{\alpha_2 t\},
\quad \quad
b_1=h_3 exp\{\alpha_3 t\}+h_4 exp\{\alpha_4 t\},
\end{equation}
where $h_{1,2,3,4}$ are constants and $\alpha_{1,3}<0$. Constants $\alpha_{2,4}>0$ on contraction stage so at late times both $b_1$ and $a_1$ diverge exponentially, whereas on expansion stage $\alpha_{2,4}<0$. One can object to this conclusion as equations (\ref{perteqsser}) are valid only for short time intervals. Comparison to numerical results shows however that perturbations diverge indeed and approximate solutions of (\ref{perteqsser}) correctly illustrate the situation.

Finally, it is necessary to note that the coefficient $C_0$ grows exponentially with $s$ and it is positive for $s>0$. Therefore, the role of the term $\psi_1^2$ becomes crucial far from the bounce on contraction stage.

\section{Numerical results}

Our analysis of the dynamics of perturbations results in a strong conclusion, namely that perturbations of the Hubble parameter and the scale factor diverge exponentially with time on contraction stage. The same happens with scalar field derivative perturbations at contraction stage while after Hubble parameter becomes positive all functions $a_1,\,b_1$ and $\psi_1$ converge exponentially with time and therefore are asymptotically stable.

\begin{figure}[ht]
\begin{center}
\epsfxsize=4.5in
\epsfbox{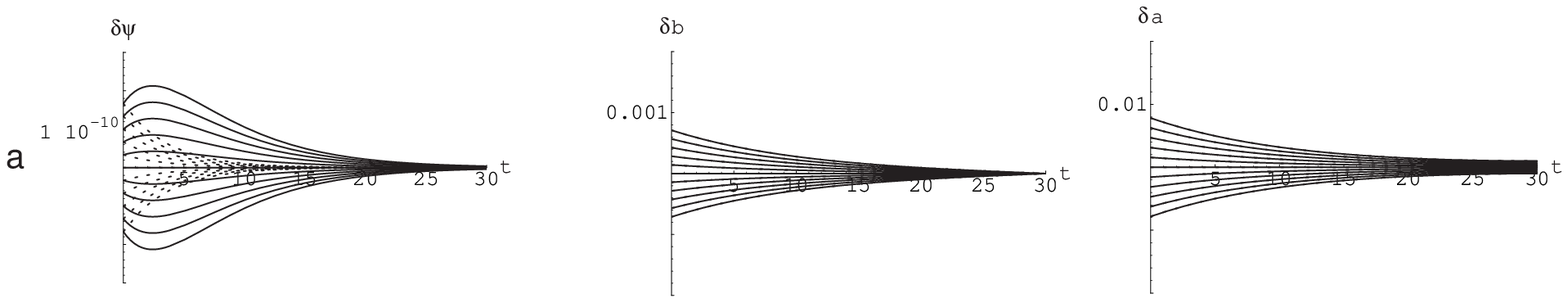}
\epsfxsize=4.5in
\epsfbox{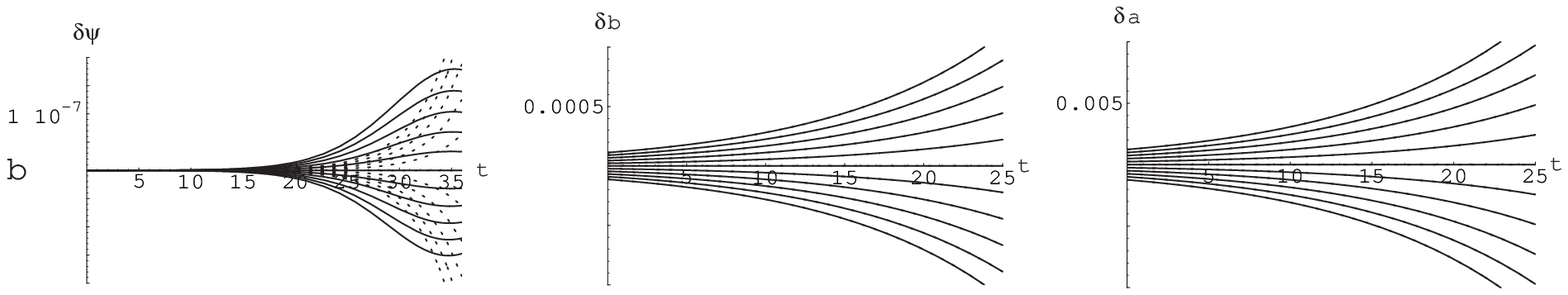}
\epsfxsize=4.5in
\epsfbox{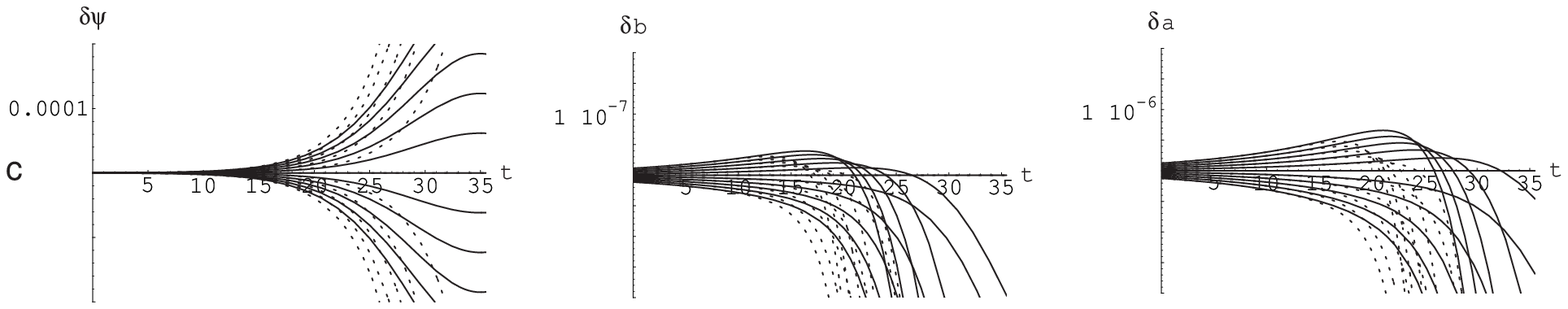}
\end{center}
\caption{Perturbations dynamics at different stages is shown. Dotted curves are exact solutions of approximate equations (\ref{perteqsser}), whereas firm curves represent numerical solutions of exact perturbative equations (\ref{perteqs}). From top to bottom figures show perturbations at de Sitter expansion stage or inflation (a) de Sitter contraction stage (b) and the same stage but with large $\psi_1$ (c).}
\label{fig2}
\end{figure}

Analytical expressions for perturbations are, however, solutions of extremely simplified perturbative equations, where time dependence in corresponding coefficients is neglected. It is interesting to compare our results with numerical solutions of full equations (\ref{perteqs}). We provide this comparison at fig.\ref{fig2}. Since function $b_1$ is a derivative of the scale factor perturbation $a_1$ it gives insight into the dynamics of the Hubble parameter perturbations.

In effect, since $b_0$ and $\psi_0$ cross zero at the bounce this will cause problems in understanding dynamics of perturbations in $b_1/b_0$ representation. So functions $b_1,\,a_1$ and $\psi_1$ are shown at corresponding figures but not dimensionless quantities $b_1/b_0$ etc.

We have chosen the value of the scalar field potential as $V_0=10^{-3}M_p^4$ so the transition from contraction to expansion takes about $t_{tr}\simeq 20t_p$. Perturbations are calculated for $s=-30$ at fig. \ref{fig2}a when transition is already finished so the dynamics of perturbations correspond to inflationary stage. It is clear that exact solutions mimic the dynamics for the scale factor and its derivative perturbations. On the other hand, on inflationary stage perturbations in the scalar field derivative converge but more rapidly than numerical calculations give.

Analytical solution of (\ref{perteqsser}) fails to reproduce perturbations dynamics during transition from contraction to expansion with negative Hubble parameter, as can be seen from the fig. \ref{fig2}b, where parameter $s=35$ and calculations are made until the bounce. Perturbations for the Hubble parameter and the scale factor are in good agreement with numerical calculations. On de Sitter contraction stage perturbations in $\psi_1$ diverge exponentially as suggested by (\ref{psi1}), moreover, divergence rate of (\ref{psi1}) is smaller than one obtained numerically. At the same time, the dynamics of scalar field derivative perturbations is incorrect in approximation used, comparing to numerical results during transition from contraction to expansion since corresponding perturbations diverge while numerical results suggest finite perturbations.

Interesting phenomenon occurs when perturbations in the scalar field derivative are assumed to be large. The term $C_0\psi_1^2$ in equation (\ref{perteqsser}) plays the crucial role in the dynamics of perturbations $b_1,\,a_1$. Equations (\ref{perteqsser}) can be integrated analytically in the presence of $C_0$ term. This case is shown at fig.\ref{fig2}c. Again analytical and numerical results are in good agreement except for $\psi_1$ function at transition stage. This is in fact the most interesting case because perturbations in the Hubble parameter and the scale factor diverge exponentially with time but there is an asymmetry so the Hubble parameter tends to large \emph{negative} values.

In realistic models approximation $V\approx \mbox{const}$ is satisfied during inflation and it is also valid during exponential contraction in nonsingular models. While the scalar field changes slightly, scalar field derivative grows exponentially and the kinetic part of the energy density soon starts to dominate over potential part. This is why the effective equation of state on contraction stage tends to $p=\rho$ in the case of GR and $p=\rho/3$ within GTG. Therefore, conditions (\ref{GRbounce}) cannot be satisfied for majority of solutions both within GR (with minimal or nonminimal coupling) and GTG at the bounce.

\section{Discussion}

Before going to conclusions it is useful to mention briefly the sequence of assumptions used in this paper.

\begin{itemize}
	\item In Friedmann equations we neglect term with scalar field derivative with respect to time and assume that the scalar field is approximately constant, so the effective potential is flat.
	\item In the scalar field equation (conservation law) it is assumed that the derivative of the effective potential with respect to the scalar field is nonzero and it is a small constant. In zero order this results in good approximation for the derivative of the scalar field dynamics comparing to the case of realistic effective potentials.
	\item Perturbed equations are not linearized and solved numerically. Results agree with differences of close solutions with the same initial conditions, as for perturbations.
	\item Time-dependent coefficients in linearized perturbed equations are expanded into series keeping only zero-order terms so coefficients lose dependence on time. These approximate equations are integrated analytically.
	\item Approximate linearized perturbed equations with quadratic on scalar field derivative term are integrated analytically in the case of large perturbations of the scalar field derivative. The result is in good agreement with numerical calculations.
\end{itemize}

The results suggest that nonsingular solutions with chaotic inflation are exponentially unstable on contraction stage and therefore are of little interest. In effect, our conclusion supports one obtained long time ago in \cite{Star,BGZH} on the base of different approaches.

Behavior of flat inflationary models with massive scalar field within GTG is analysed in \cite{Ver03} where it is pointed out that bouncing models have zero measure like in the case of GR. Results of this paper support conclusion about instability of such models.

\section{Conclusions}

Perturbative analysis of nonsingular inflationary cosmological equations of GR (in the case of minimal as well as nonminimal coupling) and GTG is given. In approximation of slowly varying scalar field cosmological equations can be reduced to the case of massless scalar field and cosmological constant. Approximate equations for perturbations with time-independent coefficients are solved analytically and compared to numerical solutions of exact perturbative equations.

It is shown, that perturbations of the Hubble parameter, scale factor and the scalar field derivative are asymptotically (exponentially) stable during inflation. On the contrary, on exponential contraction stage all perturbations are unstable and diverge exponentially. For large perturbations of the scalar field derivative and far from the bounce perturbations in the Hubble parameter tend to large negative values while for small perturbations of the scalar field derivative the Hubble parameter tends equally to large positive and negative values.

\end{document}